# Gravitation and the earth sciences: the contributions of Robert Dicke

Helge Kragh[*]

**Abstract:** The American physicist Robert Dicke (1916-1997) is primarily known for his important contributions to gravitation, cosmology, and microwave physics. Much less known is his work in geophysics and related areas of the earth sciences in which he engaged himself and several of his collaborators in the period from about 1957 to 1969. Much of Dicke's work in geophysics was motivated by his wish to obtain evidence in support of the non-Einstenian Brans-Dicke theory of gravitation. The idea of a decreasing gravitational constant, as entertained by Dicke and some other physicists (including Pascual Jordan), played some role in the process that transformed the static picture of the Earth to a dynamical picture. It is not by accident that Dicke appears as a minor actor in histories of the plate tectonic revolution in the 1960s.

## 1. Introduction

Although not a recipient of the Nobel Prize, Robert Henry Dicke (1916-1997) is recognized as one of the most eminent, influential, and versatile physicists in the second half of the twentieth century [Happer, Peebles, and Wilkinson 1999; Peebles 2008]. Equally at home in experiment and theory, Dicke straddled the barrier that still in the 1950s made quantum mechanics and general relativity two separate worlds. In the context of American post-World War II physics he was also unusual by his deep interest in foundational and philosophical aspects of physics, including the classical discussions of Isaac Newton, George Berkeley, and Ernst Mach. Dicke was particularly fascinated by Mach's principle according to which (in one of its several versions) the space-time metric is determined by the distribution of mass in the universe. He was fascinated, even obsessed by the principle of the Viennese philosopher-physicist, which was the philosophical beacon guiding much of his research concerning gravitation and cosmology [e.g. Dicke 1964a]. While most American physicists of his generation considered history and philosophy of science an unnecessary luxury, to Dicke it was an inspiring force. During his distinguished career he worked successfully in a broad range of science, which apart from quantum theory and gravitation physics also included geophysics, astronomy, and space science. Today he may be best known for his pioneering contributions to cosmology.

---

[*] Department of Mathematics, Aarhus University, 8000 Aarhus, Denmark. Address after 1 March 2015: Niels Bohr Archive, Niels Bohr Institute, 2100 Copenhagen, Denmark. E-mail: helge.kragh@css.au.dk.

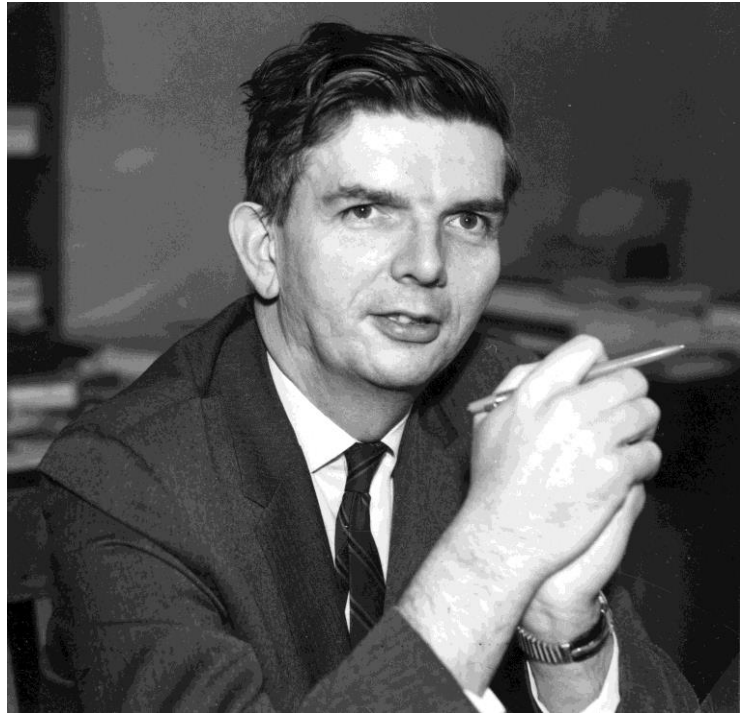

Fig. 1. Robert Dicke. Courtesy American Institute of Physics.

When young Dicke arrived at Princeton University in 1946 he focused on quantum optics, measurements of atomic structure, and the interaction of matter and radiation. Together with his Princeton colleague and former student James Wittke he published an excellent textbook in quantum mechanics, the introduction of which – dealing with "the nature of physical theories and the scope of mechanics" – demonstrates his philosophical inclination [Dicke and Wittke 1960]. Dicke was an authority in microwave radar, a branch of applied physics he had first encountered during World War II while working at MIT's famous Radiation Laboratory. His later research in experimental radiation physics proved important for the invention of the laser [Bromberg 1991]. Dicke held fifty patents, some of them on laser technology.

From the mid-1950s onwards Dicke increasingly turned from radiation and quantum physics to studies of gravitational physics, both experimentally and theoretically. He soon became a leading figure in what is known as the "renaissance of general relativity" [Kaiser 1998]. Dicke's interest in gravitation theory and cosmology, including aspects of geophysics, was first developed during a sabbatical year spent at Harvard in 1954-1955. Since there was nobody at Harvard to discuss the subjects with, he studied them on his own. In an interview with astrophysicist Martin Harwit [AIP 1985] he recalled:

> I got interested in astrophysics and geophysics because I thought these subjects provided a tool for getting at some of the questions which my interest in general relativity were bringing up. I was interested early in a number of aspects of relativity, and one of these came about from Dirac's arguments about the large numbers. Another was the interest I had in Mach's principle, thinking this should be significant. … In that framework you have the requirement that the gravitational constant is not a real constant but it's a function of coordinates. The cosmological solution varies with time, which carries with it obvious geophysical and astrophysical implications, if true.

A few years later Dicke participated in one of the important events of the renaissance of general relativity, the 1957 Chapel Hill conference held at the University of North Carolina. At about the same time that Dicke became seriously interested in gravitation and cosmology he began to explore the geophysical consequences of a varying gravitational constant $G$ such as implied by the Brans-Dicke theory of relativity. The result was a series of publications on geophysics the ultimate aim of which was to test the varying-$G$ hypothesis. The papers appeared not only in the traditional physics literature but also in journals devoted to geophysics and the earth sciences such as *Journal of Geophysical Research*. In a book-length chapter on "Experimental Relativity" from 1964 Dicke reprinted many of his early papers on relativity, cosmology, and geophysics [Dicke 1964d, pp. 165-316].

During the same period of intense work Dicke and his group in Princeton became increasingly involved in cosmology, culminating in a seminal paper of 1965 in which the Princeton group effectively founded the new hot big bang cosmology based on an analysis of the recently discovered cosmic microwave radiation. The period in question was not only one in which a new picture of the universe emerged, it was also the period of the plate tectonic revolution resulting in a new picture of the Earth. Dicke was involved in both revolutions, if more centrally in the first than the latter. This essay is primarily concerned with his lesser known contributions to geophysics in the years between 1957 and 1969. These contributions focused on the geophysical consequences of a varying gravitational constant, a hypothesis that Dicke came to independently but had previously been proposed by Paul Dirac. For this reason I start with a brief review of Dirac's hypothesis.

## 2. The Large Numbers Hypothesis

According to Dirac's so-called Large Numbers Hypothesis (LNH), when two very large pure numbers of the order $10^{39}$ and $10^{78}$ – or generally $(10^{39})^n$, where $n$ is a natural number – occur in nature, they must be connected by a simple mathematical relation [Dirac 1937; Dirac 1938]. Although generally known as the LNH, when Dirac introduced it he referred to the "Fundamental Principle." He only coined the term Large Numbers Hypothesis in 1972. Dirac considered three very large dimensionless constants of nature, namely

$$\frac{T_0}{e^2/mc^3} \cong 2 \times 10^{39}$$

$$\frac{e^2}{GmM} \cong 7 \times 10^{38}$$

$$\frac{\rho(c/H_0)^3}{M} \cong 10^{78}$$

$T_0$ is the present Hubble time, $T_0 = 1/H_0$ with $H_0$ the Hubble parameter, and $\rho$ is the mean density of matter; the symbols $m$ and $M$ refer to the mass of an electron and a proton, respectively. The first equation is a measure of the age of the universe in terms of atomic time units. From this equation and the second one Dirac suggested that $G$ varies with time as

$$G \sim \frac{1}{t} \quad \text{or} \quad \frac{dG}{Gdt} \sim -\frac{1}{t}$$

From the first and third of the equations he similarly suggested that the number $N$ of protons (or nucleons) in the universe would increase with the square of cosmic time, $N \sim t^2$. However, when he developed his idea into a cosmological model [Dirac 1938], he decided to abandon $N \sim t^2$ and only keep the $G(t)$ hypothesis. Dirac's cosmological model implied two consequences both of which were problematic. First, $G$ should vary as

$$\left(\frac{1}{G}\frac{dG}{dt}\right)_0 = -3H_0 \cong -10^{10} \text{ yr}^{-1}$$

Second, the age of the universe would be embarrassingly small, namely smaller than the age of the Earth,

$$t_0 = \frac{T_0}{3} \cong 7 \times 10^8 \text{ yr}$$

Dirac's LNH and cosmological model attracted little scientific interest. Until 1947 his 1938 paper was only cited six times and until 1957 the total number of citations was no more than twenty (Web of Science).

## 3. Brans-Dicke gravitation theory

In Einstein's general theory of relativity gravitation is fully described by the geometry of space-time as given by the metrical tensor $g_{\mu\nu}$. According to the class of "scalar-tensor" theories developed in the 1950s and 1960s a scalar field φ must be added to the tensor equations to account for the measurable value of the gravitational constant. The value of the new φ field, and hence the gravitational constant, depends on the point in space-time, φ = φ($x, y, z, t$).

The scalar-tensor formalism became well known only after Dicke and his former Ph.D. student Carl Brans published a paper on it in 1961, but most of the formalism had been established earlier by Pascual Jordan and his collaborators in Hamburg. In fact, elements of the scalar-tensor formalism can be found even earlier [Goenner 2012]. The kind of theory is often known under the names of Brans and Dicke alone, but Jordan-Brans-Dicke is also used. Brans [2014] has noted that "if a list of names were to be used for people who independently proposed ST [scalar-tensor] modifications of standard Einstein theory, the resulting compound title would be extravagantly unwieldy." What matters in the present context is that theories of this class include a gravitational constant that varies in time.

The paper by Brans and Dicke was an outgrowth of Brans' Ph.D. thesis completed earlier the same year [Brans 1961; Brans and Dicke 1961]. Although at first the new gravitation theory did not make much of an impact, within a few years it became recognized as an interesting alternative to standard general relativity [Kaiser 1998]. Until 1981 it had been cited in about 500 articles and today the cumulative number of citations is close to 2,700. The Brans-Dicke theory was based to a large extent on two principles of a general nature, Mach's principle and Dirac's Large Number Hypothesis. In agreement with an earlier idea of Dicke [1959a] the two authors took the first principle to imply that the mass $M$ of the visible universe was related to its space curvature radius $R$ by

$$\frac{GM}{Rc^2} \cong 1 \quad \text{or} \quad \frac{1}{G} \cong M/Rc^2$$

Thus, as the universe expands and $M$ and $R$ change, $G$ changes accordingly. A relation of this kind can also be derived from ordinary Friedmann cosmology, but Brans and Dicke looked upon it as the outcome of individual contributions from celestial bodies at various distances. This alone indicates that $G$ is not a fundamental constant, but a quantity which depends on the large-scale structure of the universe.

To construct a new theory of gravitation "which is more satisfactory from the standpoint of Mach's principle than general relativity" Brans and Dicke replaced the gravitational constant $G$ with a long-range scalar field $\varphi$ ($x$, $y$, $z$, $t$) generated by the matter in the universe. The two quantities were related by $G \sim \varphi^{-1}$. However, the measured value of $G$ also depended on a dimensionless parameter, $\omega$, the value of which was not given by theory but could only be determined by observation. Dicke [1962a] characterized $\omega$ as a measure of the fraction of the gravitational force caused by the scalar field. In the limit $\omega \to \infty$ the equations of the Brans-Jordan theory passed asymptotically over into those of the ordinary Einstein theory of relativity. As in the theories of Dirac and Jordan, gravitation would decrease in time, but at a rate that depended on the value of the $\omega$ parameter. For example, in the special case of a flat space with vanishing pressure (a "dust model"), the rate was given by

$$G = G_0 \left(\frac{t}{T_0}\right)^{-\delta}$$

Here, $\delta = 2/(4 + 3\omega)$ and $T_0$ is the age of the universe. The subscript 0 refers to the present era. The variation could also be described as

$$\left(\frac{1}{G}\frac{dG}{dt}\right)_0 = -\frac{H_0}{1 + \omega}$$

For comparison, the rate of decrease in the Dirac-Jordan theory was $-3H_0$. "The resulting rate of decrease of the gravitational constant is 1 part in $10^{11}$ parts per year," Dicke [1962a, p. 657] wrote. "With a closed universe this rate of decrease could be as great as 3 parts in $10^{11}$ per year." In the case of $\omega = 6$, the rate is about twenty times less than according to Dirac's theory. Dicke further stated the relationship between the age of the universe and the Hubble time $T_H$ to be

$$T_0 = \frac{2 + 2\omega}{4 + 3\omega} T_H$$

In the case of ω = 6 this amounts to a value of $T_0/T_H$ that is only marginally smaller than the ratio 2/3 in the Einstein-de Sitter cosmological model. Applying their theory to the anomalous precession of Mercury's perihelion, Brans and Dicke found a value for the perihelion shift that was smaller than Einstein's by a factor given by

$$\frac{4 + 3\omega}{6 + 3\omega}$$

To secure agreement with observations they concluded that ω ≥ 6 and possibly ω ≅ 6, which was their preferred value.

Assuming that "the universe expands from a highly condensed state," Brans and Dicke also examined various cosmological models modified by the assumption of a slowly decreasing gravitational constant. In the standard Friedmann theory governed by general relativity, the expansion of the universe of average mass density $\rho$ and zero pressure follows the expression

$$\left(\frac{\dot{R}}{R}\right)^2 + \frac{k}{R^2} = \frac{8\pi G\rho}{3}$$

The speed of light $c$ is taken to be unity, $k$ is the curvature parameter, and Λ = 0. In the Brans-Dicke theory the analogous expression is

$$\left(\frac{\dot{R}}{R}\right)^2 + \frac{k}{R^2} = \frac{8\pi\rho}{3\varphi} - \frac{\dot{\varphi}}{\varphi}\frac{\dot{R}}{R} + \frac{\omega}{6}\left(\frac{\dot{\varphi}}{\varphi}\right)^2$$

With φ = constant = $1/G$ the Friedmann equation is regained. Brans and Dicke pointed out that for ω ≥ 6 their model for a flat-space universe was practically indistinguishable from the Einstein-de Sitter model based on ordinary general relativity.

In another paper Dicke [1962b] investigated in detail the cosmological and astrophysical consequences of the Brans-Dicke theory, concluding that it was the only theory able to explain the disagreement between the age of the universe and various evolutionary ages of stars and galaxies. On the other hand, on Dirac's assumption of $G \sim 1/t$ the age of the Sun and the other main sequence stars would be suspiciously smaller than the usually accepted ages [Dicke 1961b, p. 100].

The Chapel Hill conference between 18 and 23 January 1957 was an important event in the renaissance of general relativity.[1] Organized on the initiative of Bryce DeWitt and his physicist wife Cécile DeWitt-Morette its theme was broad – the role of gravitation in physics [DeWitt and Rickles 2011; Bergmann 1957]. Among the participants were John Wheeler, Peter G. Bergmann, Hermann Bondi, Thomas Gold, Robert Dicke, Felix Pirani, Stanley Deser, and Richard Feynman. The focus of the conference was on aspects of general relativity, including gravitational waves and problems of quantum gravity, but it also included a session on cosmology.

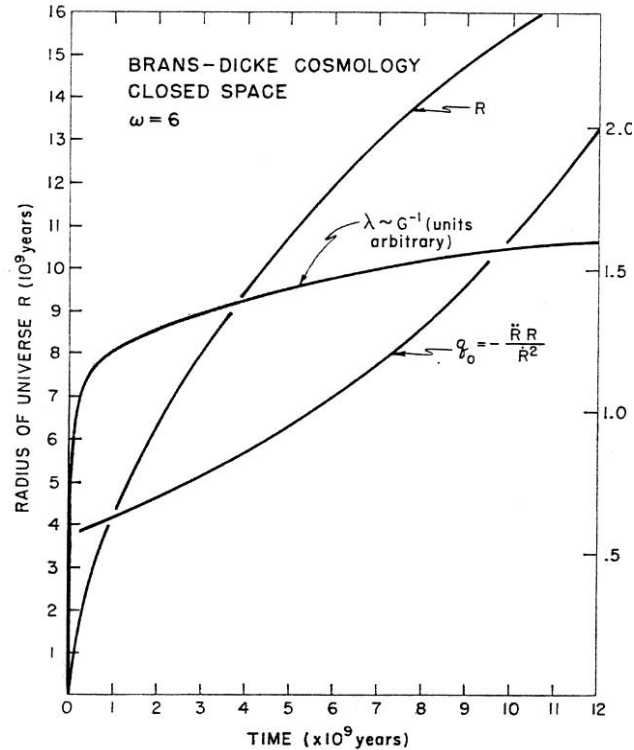

Fig. 2. The Brans-Dicke universe for ω = 6. *R* is the scale factor and $q_0$ the curvature parameter. Source: Dicke 1962a, p. 636.

Dicke's talk at Chapel Hill on "The Experimental Basis of Einstein's Theory" mostly dealt with the consequences of a possible decrease in time of the gravitational constant. What was the meaning of the "famous dimensionless numbers," he wondered. He listed three possible answers [Dicke 2011, p. 53]:

> First, and what ninety percent of physicists probably believe, is that it is all accidental; approximations have been made anyway, irregularities smoothed out, and there is really nothing to explain; nature is capricious. Second, we have Eddington's view, which I may describe by saying that if we make the mathematics complicated enough, we can expect to make things fit. Third, there is the view of Dirac and others, that this pattern indicates some connections not understood as yet. On this view, there is really only *one* "accidental" number, namely, the age of the universe; all the others derive from it. The last of these appeals to me; but we see immediately that this explanation gets into trouble with relativity theory, because it would imply that the gravitational coupling constant varies with time. Hence it might also well vary with position.

---

[1] Another major conference on general relativity, celebrating the fiftieth anniversary of Einstein's relativity theory, was held in Bern in 1955 [Mercier and Kervaire 1956]. While Jordan did not participate in the Chapel Hill conference, in Bern he gave a lecture on mathematical aspects of his varying-*G* gravitation theory. Several of the participants in Bern would also come to Chapel Hill two years later, including Bergmann, Deser, Bondi, and Pirani. Whereas the Bern conference had been attended by only nine US-based physicists out of a total of ninety or so, Chapel Hill was dominated by the US relativity community. It was smaller and more oriented towards recent work in general relativity than the Swiss conference.

Like Jordan in Germany, Dicke was fascinated by Dirac's idea but in a more critical and independent way. From Dicke's point of view, the numbers of orders $10^{39}$ and $10^{78}$ could not have been much different, since they were conditioned by the presence of intelligent life [Dicke 1957a, p. 356; Dicke 1957b, p. 375]. Humans could not have evolved had the age of the universe $T$ been much smaller than $10^{39}$ atomic time units, nor would they exist if the age was much greater. Over the next several years he amplified his argument that Dirac's reasoning contained a "logical loophole" by assuming the epoch of humans to be random. "With the assumption of an evolutionary universe, $T$ is not permitted to take one of an enormous range of values, but is somehow limited by the biological requirements to be met during the epoch of man," he wrote [Dicke 1961a; Dicke 1959a]. His paper in *Nature* caused a brief reply from Dirac, who now, for the first time since 1938, returned to cosmology. Dicke's biologically oriented interpretation of the LNH eventually became an important stimulus to the anthropic principle introduced by Brandon Carter in 1973 [Barrow and Tipler 1986; Kragh 2011, pp. 220-228].

In his address to the Chapel Hill conference Dicke dealt for the first time with some of the astronomical and geophysical effects of $G(t)$, such as the climate in past geological ages, the formation of the Moon, and the heat flow out of the Earth. A more formal and much more elaborated version appeared half a year later in two papers in *Reviews of Modern Physics*.[2] In one of them Dicke [1957b] proposed a rather

---

[2]  Issue no. 3 of the 1957 volume of *Reviews of Modern Physics* contained papers prepared in connection with the Chapel Hill conference. Most of the papers, including Dicke [1957b], had

speculative theory of gravitation based on an analogy to a polarizable vacuum described by a time-varying permittivity. He suggested that his new picture of empty space – or ether, as he later said [Dicke 1959a, p. 29] – might lead to the creation of particles in an originally matter-free primordial universe containing only gravitational energy. The other of the papers, more relevant to the context of this essay, was published under the rather misleading title "Principle of Equivalence and the Weak Interactions." Not only was most of the paper concerned with geological consequences of a varying gravitational constant, it also had nothing to do with weak interactions as usually understood.[3] Dicke took the term to mean interactions described by a small coupling constant, which primarily meant gravitation.

## 4. Jordan's expanding Earth

Dicke was not the first physicist to investigate the geophysical consequences of a varying $G$. He was well aware of the similarity between the Brans-Dicke gravitation theory and an earlier theory of the scalar-tensor type proposed by Jordan, and also aware that Jordan had applied the $G(t)$ hypothesis to geological phenomena. While Jordan's work on gravitation was not well known by many English-speaking physicists [Goenner 2012], Dicke was always careful in acknowledging the priority of his colleague in Germany with whom he met and corresponded [Peebles 2008]. Still when Brans had almost completed his thesis he was unaware of Jordan's theory, but (undoubtedly due to Dicke) the final form of it included an extensive discussion of the theory as presented in Jordan's book *Schwerkraft und Weltall* [Brans 2010; Jordan 1955, first edition 1952].

Jordan was an early and for long the only advocate of Dirac's gravitation hypothesis which he defended and developed in books and articles between 1937 and his death in 1980. What matters in the present context is that since the early 1950s Jordan devoted much of his research to prove that the history of the Earth

---

not been presented at the conference but may have been informally discussed. The other paper by Dicke [1957a] followed his presentation at Chapel Hill.

[3] The only reference in Dicke [1957a, p. 362] to what normally is called weak interactions was at the end of his paper, where he inferred from Dirac's LNH that "β decay rates would vary inversely as the square root of the age of the Universe." The same result had earlier been suggested by Jordan which Dicke may not have been aware of. A few years later he considered the possible variation in time of the weak coupling constant in relation to the age of meteorites [Dicke 1959b; Peebles and Dicke 1962b]. He suspected the beta decay rate to vary as $t^{-n}$, with $\frac{1}{4} < n < \frac{1}{2}$, but available evidence was too uncertain to show any variation.

provided convincing evidence for the $G(t)$ hypothesis.[4] The main result of a decreasing $G$, according to Jordan, was that the Earth is expanding, thereby offering an explanation of the separation of the continents without accepting Alfred Wegener's controversial idea of continental drift. He also argued that the $G(t)$ hypothesis could explain the formation of mountains, volcanism, and paleoclimatic phenomena such as the ice ages. In a monograph of 1966, translated five years later into English, he presented his arguments in a systematic and comprehensive form [Jordan 1971].

In contrast to Dicke, Jordan worked mostly alone and without seeking advice from or collaboration with earth scientists. He seems to have rated the opinions of geologists and geophysicists lowly. At a symposium in honour of Dirac's seventieth birthday he said, "I came to the decision not to ask other specialists [in the earth sciences] about the compatibility of Dirac's hypothesis with empirical facts, but to try to learn myself what really are the proven facts, and to see whether they lead to real contradictions against Dirac's hypothesis" [Jordan 1973, p. 61]. Although Jordan realized that the evidence in favour of Dirac's hypothesis was indirect and far from generally accepted, he was convinced that the evidence from geology and astronomy proved the correctness of the hypothesis. Since the expansion of the Earth followed from the $G(t)$ hypothesis, this was also to be regarded a proven fact. In a paper in *Reviews of Modern Physics* Jordan maintained that "our present knowledge of the earth … makes the correctness of Dirac's hypothesis an established fact" [Jordan 1962, p. 600].

## 5. The turn towards geophysics

As is evident from his Chapel Hill address and the two subsequent papers, by 1957 Dicke had obtained solid knowledge of a series of geophysical subjects, which he discussed expertly. This was a new field to him. While an undergraduate in Princeton in the late 1930s he had the admirable intention of taking at least one course in each of the major sciences. This he did with only one exception, namely geology [AIP 1985]. But latest by 1957 he had become seriously interested in the subject, as indicated by his membership of the American Geophysical Union.

Dicke learned much of his geophysics and geology by self-studies, but he also acknowledged "the many suggestions and ideas I have derived from conversations with Professor H. Hess of the Princeton geology department" [Dicke 1962a, p. 664; Dicke 1961b, p. 106]. He was referring to Harry Hammond Hess, who had joined the faculty in 1934 and in 1950 was made head of the Department of Geology. Ten years later Hess formulated the crucial idea of sea floor spreading and thereby made a seminal contribution to what would soon become known as plate tectonics [Frankel 2012, pp. 280-319]. Dicke [AIP 1975] recalled:

> Long before the average geologist in the country took this continental drift, and plate tectonics, to mean anything at all – Harry Hess, over in our geology department, had a clear picture of what was going on. Anyway, it's obvious that as gravitation is getting weaker, the earth should expand slightly, and I noticed in my readings at that time that there were cracks in the mid-Atlantic ridge in the ocean, oceanic cracks. So this suggested that these cracks might be the result of tension, due to the earth expanding … . I went over and talked to Hess about this. We laid out a beautiful picture of the Atlantic Ocean crust moving, and trenches, and island arcs, and all the – the whole plate tectonic game was laid out for me, and this was the late fifties.

---

[4]  For details on Jordan's work in geophysics, see H. Kragh, "Pascual Jordan, varying gravity, and the expanding Earth," submitted to *Physics in Perspective*.

In addition to their shared interest in geophysics, Hess and Dicke also had in common their involvement in the new space sciences. Until his death in 1969 Hess served as chairman of the Space Science Board established under the National Academy of Sciences in 1958.

On the suggestion of Hess, Dicke visited Yale in 1959 to discuss with the Australian geologist Samuel Warren Carey the evidence for an expanding Earth and its relation to decreasing gravity. Carey, who stayed as visiting professor at Yale, had since 1956 advocated the view that the present surface of Earth was the result of a long phase of expansion. He argued that the Earth's radius had increased since the early Paleozoic era some 500 million years ago at an average rate of 5 mm per year. In late 1959 and early 1960 Carey gave several lectures in Princeton, where he presumably met Dicke [Carey 1988, p. 119 and p. 141].

From Jordan's book *Schwerkraft und Weltall*  Dicke was already familiar with the idea of Earth expansion caused by a decreasing gravitational constant. Generally Dicke had easy access to several of the American geophysicists and oceanographers who were instrumental in the plate tectonic revolution. Moreover, at Princeton he mobilized several physics graduate students to write theses on geophysical and astrophysical problems related to the $G(t)$ hypothesis.

William Jason Morgan wrote in 1964 his physics Ph.D. thesis under Dicke and with Hess on his committee. The title of Morgan's thesis was "An Astronomical and Geophysical Search for Scalar Gravitational waves." Such "φ-waves" were expected from the Brans-Dicke theory and Dicke [1964c] wanted to know if they actually existed and what observable effects they might have. One possibility was that the φ-waves, which were caused by a decrease in $G$, triggered earthquakes. This was the subject that 26-year-old Morgan investigated in a paper co-authored by J. O. Stoner and Dicke [1961]. By calculating the correlation between earthquake rates and changes in the Earth's rotation he found a correlation value that might possibly be ascribed to a changing $G$. At the time Morgan had taken no course in geology or geophysics, but his work with the thesis turned him into a geophysicist. At the end of 1964 he was hired by the German-American geophysicist Walter Elsasser, who had joined Hess in Princeton in 1963. Four years later, in what was only his fifth paper, Morgan presented a cornerstone of the new plate tectonics in the form of a quantitative, mathematically formulated theory of trenches and faults.

When Morgan received the prestigious National Medal of Science in 2003 he noted that Dicke had received the same award in 1970. He paid the following tribute to his former thesis adviser [Schultz 2003]: "My apprenticeship with him more than 40 years ago was where I learned what science is – how to formulate and attack a scientific problem. His approach and attitude toward science remain with me today."

## 6. Varying gravity and geophysics

In broad-ranging papers around 1960 Dicke surveyed how a varying gravitational constant as given by either Dirac's theory or the slower version of the scalar-tensor theory would affect the Earth and the Moon. His aim was not primarily to contribute to the geological sciences, but rather to use geological data as tests for the $G(t)$ hypothesis and more specifically for the Brans-Dicke theory of gravitation. This was basically the same aim that Jordan pursued. However, not only were Dicke's investigations generally of a more detailed and quantitative nature, his conclusions were also more reserved than Jordan's. None of the available evidence "can be used to give strong support to Dirac's hypothesis," Dicke cautiously stated in his 1957 paper, adding that a variation of $G$ could not be excluded. Five years later, he concluded that his analysis of examples from geophysics "cannot be marshaled as evidence for a gradual decrease in the gravitational constant" [Dicke 1962a, p. 664; Dicke 1964b, p. 173]. To him, the case for $G(t)$ was neither proven nor disproven. The two physicists also had different attitudes to the expanding Earth, which Jordan was strongly committed to. Dicke, on the other hand, only supported the hypothesis half-heartedly and for a rather brief period of time.

In 1948 Hungarian-born nuclear physicist Edward Teller had argued that Dirac's $G(t)$ hypothesis led to consequences in conflict with established paleontological knowledge [Teller 1948]. Assuming that the absolute temperature of the Earth's surface $\theta$ depends directly on the energy flux received from the Sun, he found it to vary as

$$\theta \sim G^{9/4} M_S^{7/4} \,,$$

where $M_S$ is the mass of the Sun. Further assuming $M_s$ = constant it follows from Dirac's $G \sim 1/t$ that

$$\theta = \theta_0 \left(\frac{t_0}{t}\right)^{9/4}$$

The symbol $t_0$ denotes the present cosmic epoch or roughly the Hubble time. Teller concluded that at a time 200-300 million years ago the surface temperature of the Earth would have been near the boiling point of water. Since palaeontology showed "ample evidence of life on our planet at this time," he concluded that Dirac's hypothesis was probably wrong. Teller's argument was often cited as proof of the incorrectness of the $G(t)$ hypothesis, but as Dirk ter Haar [1950, p. 131] pointed out it was actually inconclusive. Not only did it rest on the assumption of a constant solar opacity, it also disregarded cloud formation in the atmosphere. If heavy clouds were formed all over the Earth, the average temperature would be considerably lowered.

Dicke reconsidered in detail the question of the Earth's temperature in the past. Whereas Teller had assumed a Hubble time of about $2 \times 10^9$ years, ten years later astronomers believed that the parameter might be as much as ten times larger. Dicke [1962a] assumed an age of the universe of $8 \times 10^9$ years and $G$ to vary in accordance with the Brans-Dicke theory with $\omega = 6$, meaning

$$\left(\frac{1}{G}\frac{dG}{dt}\right)_0 \cong -1.2 \times 10^{-11} \ \text{yr}^{-1}$$

On the Dirac-Jordan hypothesis of $G \sim 1/t$ the rate of decrease of $G$ would be larger, about $2.4 \times 10^{-10}$ yr$^{-1}$. Dicke took into account the great amount of water vapour in the cloud-covered atmosphere caused by increased temperature. The effect of the water vapour, he suggested, would be to stabilize the temperature. As a result of his calculations he suggested that the surface temperature in the early period of the history of the Earth agreed with the existence of algae some three billion years ago. The analysis of 1962 thus confirmed his earlier conclusion that "there is no particular difficulty in accounting for life over a period of the past billion years" [Dicke 1957a, p. 358]. In a review of 1961 Dicke estimated the past surface temperature of both the Earth and the Moon on the assumption of Dirac's $G(t)$ and a Hubble constant of $H_0 =$ 80 km s$^{-1}$ Mpc$^{-1}$. He admitted that the curve for the Earth's variation in temperature "has no great reliability" [Dicke 1961b, p. 101].

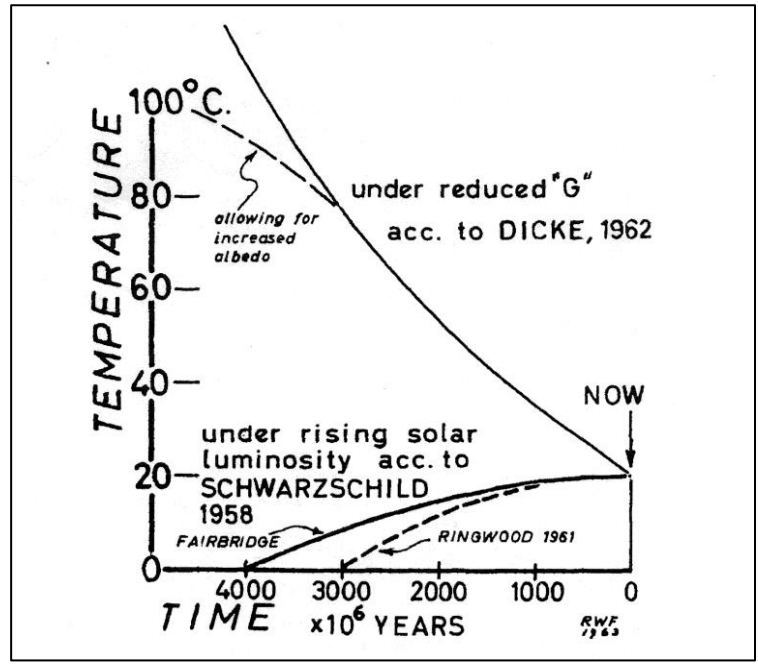

Fig. 3. Different views of the surface temperature of the past Earth. Source: Fairbridge 1964.

But was the climate of the Earth really warmer in the geological past? According to some geologists, referring to astrophysical theory and mineralogical evidence, the Precambrian had been characterized by a very *low* temperature which since then had gradually increased [Ringwood 1961; Fairbridge 1964]. This was also what Martin Schwarzschild and collaborators [1956, p. 241] had concluded from model calculations of the Sun: "In the early pre-Cambrian era, two billion years ago, the solar luminosity was about 20 per cent less than now. The average temperature on the earth's surface must then have been just about at the freezing point of water, if we assume that it changes proportionally to the fourth root of solar luminosity. Would such a low average temperature have been too cool for the algae known to have lived at that time?" Of course, this climate scenario was diametrically opposite to the one proposed by Teller, Dicke and others on the basis of a decreasing $G$.

By the late 1970s climate models indicated that the solar constant had been remarkably constant over a period of $3 \times 10^9$ years and that the change in solar luminosity during the last $3 \times 10^8$ years had been less than 3 per cent. Most experts now agreed that the brightness of the Sun was of relatively little importance for variations in the past climate of the Earth [Wigley 1981]. Questions of paleoclimatology were too messy and complex to be answered by the traditional methods of physicists and astronomers. Nor was paleoclimatology of any real use in testing competing cosmological models. The conclusion of Dicke [1964b, p. 160] was agnostic: "The moral is that the atmosphere is complicated … We cannot be sure how much the surface temperature would have changed."

### 7. Structure and heat economy of the Earth

Dicke [1957a; 1961b; 1964b] also discussed the consequences that the $G(t)$ hypothesis would have for the formation and surface of the Moon. If the Moon had originally been formed in a molten state and subsequently solidified – Dicke assumed it to have occurred 3.25 billion years ago – its surface area would since then have increased slightly because of the expansion caused by the decreasing gravity. As possible evidence for the expansion he pointed to the cracks or so-called rills on the Moon's surface. The problem of the formation of the Moon was at the time unresolved, with the capture hypothesis and the rival fission hypothesis being the most popular candidates [Brush 1996]. Dicke argued that the hotter Sun in the past, as predicted by the $G(t)$ hypothesis, supported the view that the Earth and the Moon were formed as molten bodies. He was confident that the highest temperature of the Moon would have been some 250 °C at the time of its formation. Tidal waves in the liquid Earth might have produced the mechanism for fission in accordance with the idea originally suggested by George Darwin in the late nineteenth century.

"I had one student, for example, look at the heat flow problem from this point of view [decreasing gravity], because … if you have the interior hot, as we understand it is, and if the temperature curve for the mantle of the earth is near the melting point, if you lower the pressure inside, … you can calculate the way in which heat flows out this way. And it agrees rather well with what is observed, to a factor of two." This is how Dicke recalled about his student Charles T. Murphy who wrote an undergraduate thesis on the problem [AIP 1975]. Dicke [1957a, p. 361] had earlier discussed it in a more preliminary way, noting how remarkable it was that "the rate of heat flow from the earth should present a crucial test for a physical hypothesis." Some years later an extended version of Murphy's thesis was transformed into a joint paper published by the American Philosophical Society [Murphy and Dicke 1964].

From considerations of the energy sources of the Earth Dicke and his young co-author inferred that radioactivity and other known processes were not enough to account for the observed heat production of the order 50 erg s$^{-1}$ cm$^{-2}$ reaching the Earth's surface area. The additional energy source, he suggested, might have its roots in gravity weakening in time. The idea was, roughly, that as $G$ decreases, the pressure and melting point of the mantle will decrease and the interior of the Earth cool. Heat will flow out of the core and mantle at a rate greater than if $G$ were constant. Moreover, the induced heat flow would add to the heat generated by radioactivity and make it possible for convection currents to occur in the mantle. The net result of a series of complex mechanisms was this: "The decreasing gravitational constant makes available the internal heat of the earth as a steady-state convective system in the mantle and makes convection a more likely possibility" [Murphy and Dicke 1964, p. 243]. Calculating the heat release brought about by this effect, they found the value 2.5 × 10$^{12}$ cal s$^{-1}$ for the entire Earth surface, or about 20 erg s$^{-1}$ cm$^{-2}$. "Thus a large portion of the observed heat flow might originate deep in the earth," they concluded. The proposed mechanism might also explain the "mystery" of the rate of heat flow from the ocean floors, namely, that it was nearly the same as that from the continents despite the much higher content of radioactivity in the latter. Dicke [1962a] took the mean heat flow from the ocean floor to be 35 erg s$^{-1}$ cm$^{-2}$ and found the effect of a decreasing $G$ to represent almost half this figure.

Richard von Herzen, a geophysicist at the Woods Hole Oceanographic Institution in Massachusetts, was not ready to accept Dicke's "exotic" explanation of the heat escaping from the Earth's interior. He suggested that the theory and its basis in $G(t)$ was a "speculative mechanism [that] should be accepted with a small grain of salt … until a need to accept its implications increases" [Herzen 1967, p. 213].

Murphy and Dicke were well aware that convection in the mantle was a controversial hypothesis and had remained so since the days of Wegener. Indeed, convection currents of radioactive origin were the favoured mechanism for the horizontal movements of the continents and consequently denied by Harold Jeffreys and many other critics of Wegener's drift theory. Noting the close connection between the convection hypothesis and continental drift, Jordan [1971, pp. 95-97] criticized the hypothesis as contrived and unnecessary. On the other hand, he praised the work of Murphy and Dicke for offering a new and physically more correct mechanism for mantle convection without claiming that it supported continental drift. None of the effects of a varying $G$ that Murphy and Dicke predicted for the interior of the Earth violated established geophysical facts, but unfortunately the effects were not empirically verified. Consequently, the two authors concluded, "they do not directly verify the [varying *G*] hypothesis."

## 8. Drifting ideas and drifting continents

Nor did detailed studies on the rotation of the Earth and the temperature of meteorites succeed in clearly identifying effects of the decreasing gravitational constant. Dicke [1966] concluded that the astronomical data allowed a rate of change of *G* in agreement with the Brans-Dicke theory (of the order $10^{-11}$ yr$^{-1}$), but was unable to present his analysis in stronger terms, that is, as positive evidence for the theory. He returned to the issue in an article of 1969, this time with a focus on the geophysical consequences rather than the varying constant of gravitation [Dicke 1969]. The situation was about the same in an investigation made with his former student James Peebles concerning the amount of argon in meteorites [Peebles and Dicke 1962a].

Peebles came to Princeton in 1958 to study particle physics but soon became part of Dicke's Gravity Group. On the suggestion of Dicke and motivated by his fascination with Mach's principle, Peebles wrote in 1962 his dissertation on varying constants of nature. About his encounter with Dicke and his group, Peebles recalled [Peebles, Page, and Partridge 2009, p. 185]: "Bob's motivation was his fascination with Mach's principle, which might be read to say that as the universe evolves so do the laws of physics. I was fascinated by all the evidence one could bring to bear, from the laboratory to geology and astronomy. My evident lack of interest in Mach didn't seem to bother Bob." Some of the papers co-authored by Dicke and Peebles in the 1960s relied in part on Peebles' dissertation [Peebles and Dicke 1962b; Peebles and Dicke 1962c].

The idea of the Dicke-Peebles study of meteorites was that a higher temperature in the early history of the solar system would have caused an anomalous loss of argon from meteorites. But the method only made it possible to estimate an upper value for the relative *G*-dependence on time of $10^{-10}$ per year. When Dicke and Peebles [1965] reviewed the empirical arguments for the Brans-Dicke theory, the words "not compelling" appeared repeatedly. It may have been this state of affairs, namely, the inadequacy of astronomical, geological and geophysical tests to yield unambiguous answers that caused Dicke to withdraw from geophysics. For a few more years he continued doing work in the area, but then abandoned it. According to Dicke [AIP 1975], a weakening gravity and an expansion of the Earth might play some role in terrestrial history, but "it just seemed to me to be so deeply buried in all the other things that it would be hard to separate it out, in an unambiguous way. … I decided after a while that it was just too hard to try to get fundamental physics out of the earth."

Despite all the annoying uncertainty, according to Dicke there were good reasons to believe that $G$ decreased at a rate of about $3 \times 10^{-11}$ yr$^{-1}$. If this were the case, the Earth would expand, but only at a modest rate corresponding to an increase in radius $R$ of 0.5 cm per century or 0.05 mm per year [Dicke 1957a; Dicke 1962a]. Over a period of four billion years the radius would have expanded by only 200 km. This was an expansion of a scale quite different from that advocated by Carey, according to whom the radius of the Earth in the Precambrian was only half the present radius. The relationship between $R(t)$ and $G(t)$ can in general be written as

$$\frac{1}{R}\frac{dR}{dt} = -\frac{\alpha}{G}\frac{dG}{dt},$$

where $\alpha$ is a quantity depending on the equation of state of the Earth. Dicke suggested that $\alpha \cong 1$, a value in rough agreement with later estimates. A relation of this kind was first obtained by George B. Hess, a son of the geology professor H. Hess, who in 1958 wrote an unpublished senior thesis on Dicke's Machian ideas of a varying gravitational constant.[5]

Dicke's cautious advocacy of a connection between $G(t)$ and the expanding Earth was noticed by the leading Canadian geophysicist John Tuzo Wilson, who for a brief period expressed sympathy for expansionism. Wilson [1960] thought that the radius of the Earth might have increased by approximately 0.8 mm per year, which was "close to Dicke's estimate." He similarly found the $G(t)$ hypothesis to be "an inviting idea." However, Wilson soon abandoned expansionism for the revived theory of continental drift. He is today known as one of the fathers of plate tectonics.

Could the system of oceanic ridges and the distributions of the continents be the result of the gravity-induced expansion of the Earth? Not according to Dicke, who emphasized that the required magnitude of the expansion could not be explained in terms of gravity decreasing at a rate of about $10^{-11}$ per year. The rate would have to be higher by a factor of one hundred, for which there was no physical justification. Dicke [1964b, p. 162] consequently dismissed Carey's suggestion of a rapid expansion of this magnitude. In his view, the geological evidence rather pointed towards the continental drift picture with convection in the Earth's mantle as the driving mechanism. As to the effect of a diminishing $G$, he wrote that "The miniscule effects of a modest expansion would be lost in the magnificent displays produced by convection" [Dicke 1962a; Murphy and Dicke 1964]. This was also his conclusion in a report to the Space Science Board [Dicke 1961b, p. 105]:

> It is not clear that a general expansion of the Earth is required to cause such a separation [of the Americas and Europe-Africa]. It could also be caused by convection in the mantle. Certainly it must be said that if continental drift has been occurring to the extent indicated by recent paleomagnetic data, the effects of an expansion of radius 47 km per b.y. would be negligible. … If subcrustal currents are not important, an expansion in radius of only 0.0047 cm per year could produce a medial crack in the Atlantic 2 km wide in only 13 m.y., assuming that half the expansion takes place in the Atlantic.

---

[5] The title of the thesis was "The Annual Variation of the Length of the Day as Evidence Relating to a Theory of Gravity" [cited in Morgan, Stoner, and Dicke 1961; Dicke 1962a]. After studies at Princeton, George Hess took his Ph.D. from Stanford in 1967 on experiments with liquid helium. He subsequently became a professor of physics at the University of Virginia, where he mostly worked in condensed matter physics. I am grateful to George Hess for having confirmed the authorship of the 1958 thesis (E-mail of 15 January 2015).

Dicke obviously was not an expansionist in the sense of Jordan or Carey, but neither was he strongly committed to the revived theory of drifting continents. He just considered it the best of the available pictures of the Earth and its history.

## 9. The hot big bang

Today Dicke may be best known for the important contributions that he and his group of Princeton physicists made to the revived hot big-bang cosmology in the wake of the discovery of the cosmic microwave background. His collaborator Peebles felt that Dicke should have shared the 1978 Nobel Prize with Arno Penzias and Robert Wilson [Peebles, Page, and Partridge 2009, p. 191]. At the time of the discovery Dicke was strongly attracted to the scalar-tensor theory with a decreasing gravitational constant. This theory guided much of his thinking about the early universe which in 1963 led him to consider an initial hot phase filled with blackbody radiation. To a large extent, what became the standard big-bang theory was indebted to a heterodox theory of gravitation, the Brans-Dicke theory. As far as Dicke was concerned it was also indebted to his preference for a cyclic universe and his belief that the big bang had its origin in the contraction of a preceding universe, a big crunch. He believed that a cyclic universe was "the most appealing possibility" [Dicke and Peebles 1965, p. 447; Lightman and Brawer 1990, pp. 201-213].

The importance of the Brans-Dicke theory is evident from a review paper that Dicke and Peebles submitted in early March 1965, shortly before they became aware of Penzias and Wilson's discovery of the 7.3 cm background. In this paper they dealt at length with the geophysical, astrophysical and cosmological consequences of the Brans-Dicke theory. Using the standard equations of general relativity, at the time Peebles had concluded that the intensity of the present (and still hypothetical) background radiation would have to correspond to a temperature of about 10 K to avoid excess helium production in the past. Should the temperature of the radiation turn out to be considerably less, they appealed to the decreasing $G$ of the Brans-Dicke theory. Because, with a larger $G$ in the cosmic past, "the universe would have expanded through the early phase very much faster than is implied by general relativity [and] this would reduce the time available for helium production, thus reducing the lower limit on the present radiation temperature" [Dicke and Peebles 1965, p. 451; Peebles, Lyman, and Partridge 2009, pp. 38-39].

A passage to the same effect appeared in the seminal paper in the July 1965 issue of *Astrophysical Journal* in which Dicke and his co-authors Peebles, Peter Roll and David Wilkinson analysed and interpreted the cosmic microwave background [Dicke et al. 1965, p. 418]. A few years later Dicke returned to the problem of helium production in scalar-tensor cosmology, where the scalar field contributes significantly to the expansion rate. His result was not encouraging: "For a flat universe (present density $2 \times 10^{-29}$ g cm$^{-3}$, Hubble age $10^{10}$ years), zero helium production would be expected. For the low-density case (present density $\sim 7 \times 10^{-31}$ g cm$^{-3}$) there are two possibilities: 32 per cent and 0 per cent helium" [Dicke 1968, p. 22]. From a modern perspective it may appear that considerations of $G(t)$ in the formation of the new big-bang theory were really irrelevant and unnecessary. However, they actually played a considerable role and indirectly provided a link to the kind of geophysical work that was similarly motivated. As Dicke saw it, both the Earth and the universe might serve as a testing ground for the scalar-tensor theory of gravitation that was the primary target for his research.

## 10. Conclusions

During the same period that Dicke turned towards gravitation physics and cosmology he developed an interest in geophysics, a branch of science he cultivated on and off for a decade or so. In addition to his own work in the area he also induced several of his students and assistants to take up problems of geophysics. Among the students was W. Jason Morgan, soon to become a leader of the plate tectonic revolution. Given that Dicke had no previous acquaintance with or interest in the earth sciences, he turned into a practising geophysicist with surprising ease. Although he was genuinely interested in the subject most of his work in geophysics and related areas (such as astrophysics and planetary science) was motivated by his enduring interest in a revised gravitation theory inspired by Mach's principle. The Brans-Dicke theory predicted $G$ to depend on time and it was the geological effects of $G(t)$ which principally motivated his research in the earth sciences. In the back of his mind was always the enigma of gravitation. After much hard work he was forced to admit that the project was a failure, at least in the sense that it proved impossible "to get fundamental physics out of the earth." And fundamental physics was that he aimed at.

Yet Dicke did not abandon the idea of varying gravity, he only realised that to test the hypothesis precise astronomical methods were better suited than the uncertain methods based on the earth sciences. The idea of laser ranging to probe the fundamentals of gravity was first proposed by Dicke and his collaborators even before the invention of the laser [Bender et al. 1973]. It eventually developed into the Lunar Laser Ranging project with the first retroreflectors being placed on the Moon in July 1969. Accumulated data from this project has shown that if $G$ varies in time it is at a much smaller rate than Dicke suspected [Müller and Biskupek 2007].